\documentclass[prb,aps,twocolumn,longbibliography,superscriptaddress,hidelinks]{revtex4-1}
\usepackage{hyperref}
\usepackage{rotating}
\usepackage{graphicx}
\usepackage{latexsym}
\usepackage{amsfonts}
\usepackage{soul}
\usepackage{color}
\usepackage{amsmath}
\usepackage{color}
\usepackage{lipsum}
\usepackage{dcolumn}
\usepackage{bm}
\usepackage{mathrsfs}
\usepackage{subfigure}
\usepackage{natbib}
\usepackage{bbm}
\usepackage{amstext}
\newcommand*{\g}{\text{\slshape g}}

\usepackage{changes}

\begin{document}

\title{Multiband superconductors with degenerate excitation gaps}

\author{Paulo J. F. Cavalcanti}
\affiliation{Departamento de F\'{i}sica, Universidade Federal de Pernambuco, Av. Prof. An\'{i}bal Fernandes, s/n, 50740-560, Recife-PE, Brazil}
\author{Tiago T. Saraiva}
\affiliation{National Research University Higher School of Economics, Moscow, 101000, Russia}
\author{J. Albino Aguiar}
\affiliation{Departamento de F\'{i}sica, Universidade Federal de Pernambuco, Av. Prof. An\'{i}bal Fernandes, s/n, 50740-560, Recife-PE, Brazil}
\author{A. Vagov}
\affiliation{Institute for Theoretical Physics III, University of Bayreuth, Bayreuth 95440, Germany}
\affiliation{ITMO University, St. Petersburg, 197101, Russia}
\author{M. D. Croitoru}
\affiliation{Departamento de F\'{i}sica, Universidade Federal de Pernambuco, Av. Prof. An\'{i}bal Fernandes, s/n, 50740-560, Recife-PE, Brazil}
\author{A. A. Shanenko}
\affiliation{Departamento de F\'{i}sica, Universidade Federal de Pernambuco, Av. Prof. An\'{i}bal Fernandes, s/n, 50740-560, Recife-PE, Brazil}

\date{\today}
\begin{abstract}
There is a tacit assumption that multiband superconductors are essentially the same as multigap superconductors. 
More precisely, it is usually assumed that the number of excitation gaps in the single-particle energy spectrum of a uniform superconductor determines the number of contributing bands in the corresponding superconducting model. 
Here we demonstrate that contrary to this widely accepted viewpoint, the superconducting magnetic properties are sensitive to the number of contributing bands even when the corresponding excitation gaps are degenerate and cannot be distinguished. In particular, we find that the crossover between superconductivity types I and II - the intertype regime - is strongly affected by difference between characteristic lengths of multiple contributing condensates. The reason for this is that condensates with diverse characteristic lengths coexisting in one system interfere constructively or destructively, which results in multi-condensate magnetic phenomena regardless of the presence/absence of the multigap structure in the single-particle excitation spectrum.   
\end{abstract}
\maketitle

\section{Introduction}

The concept of the multiband superconductivity was introduced in 1959~\cite{suhl,mos} as a possible explanation of a multigap fine structure observed in frequency dependent conductivity of superconducting ${\rm Pb}$ and ${\rm Hg}$, extracted from the infrared absorption spectrum~\cite{tinkh1,tinkh2}. Despite the long history of the concept, its detailed and unambiguous confirmation was obtained only in 2000's after experiments with ${\rm MgB}_2$ [see, e.g., Ref.~\onlinecite{canf} and references therein]. The observation of two well distinguished energy gaps in the excitation spectrum of ${\rm MgB}_2$~\cite{szabo,iav} ignited widespread interest in multiband superconductivity, boosting further experimental and theoretical studies. After a decade of intensive investigations, it became clear that multiple overlapping single-particle bands are present in many superconducting materials, ranging from iron-based~\cite{pagl} to organic high-$T_c$~\cite{mazzi} and even topological superconductors~\cite{deng1,deng2}. Recent first principle calculations have demonstrated that the Fermi surface of ${\rm Pb}$ comprises two Fermi sheets, confirming multiband nature of its superconducting state proposed~\cite{suhl} to explain pioneering experiments of Refs.~\onlinecite{tinkh1,tinkh2}.

It is widely assumed that a key marker for the multiband superconductivity is the appearance of multiple energy gaps in the excitation spectrum of a (bulk homogeneous) superconductor. Then, if the excitation spectrum does not exhibit the multigap structure, the superconducting properties are expected to be those of single-band materials. More generally, it is usually assumed that the number of excitation gaps determines the number of contributing bands in a superconducting model that captures the essential physics of interest. A well-known example is ${\rm MgB}_2$ which exhibits two excitation gaps associated with $\pi$ and $\sigma$ states~\cite{canf,szabo,iav}. Accordingly, theoretical models for superconductivity in ${\rm MgB}_2$ consider two contributing bands [see, e.g., Refs.~\onlinecite{canf,ashk,gol,gur,kosh,zhit,tanak1,tanak2}] despite the fact that the first principle calculations reveal~\cite{maz,an} four single-particle bands for ${\rm MgB}_2$, see also Ref.~\onlinecite{canf}. The two $\sigma$ bands have different microscopic parameters (diverse Fermi sheets) but degenerate excitation gaps and the same holds for the $\pi$ bands. A general perception is that the two-band model is sufficient to fully describe the superconducting state with the two spectral gaps.

However, there exists another approach that regards a multiband superconductor as a system governed by a set of competing characteristic lengths, see, e.g.,  Refs.~\onlinecite{tanak1,tanak2,nakai,babaev,saraiva}. As is well known, such a competition can lead to non-trivial physical consequences, e.g., to the spontaneous pattern formation~\cite{seul2}. Examples of systems with spontaneous patterns are well-known in the literature and include magnetic films~\cite{seul1}, liquid crystals~\cite{mac}, multilayer soft tissues~\cite{seul2, stoop}, lipid monolayers~\cite{kell}, granular media~\cite{aran} etc. A possibility of symmetry breaking patterns of vortices (labyrinth and stripes) induced by the presence of two condensate components with significantly different coherence lengths has recently attracted much interest in the context of unusual mixed (Shubnikov) phase configurations in ${\rm MgB}_2$ [see, e.g., Refs.~\onlinecite{moshch, bend} and references therein]. Coupled condensates coexisting in one material with diverse coherence lengths can interfere (interact) constructively or destructively, giving rise to phenomena absent in superconductors with a single condensate. In addition to the labyrinths and stripes of vortices mentioned above, other effects can be listed, e.g., possible fractional vortices~\cite{clecio,lin,lin1,silva1,lin2}, chiral solitons~\cite{tan1,tan2}, a giant paramagnetic Meissner effect~\cite{silva2}, enhancement of the intertype superconductivity~\cite{extGL1, extGL2}, hidden criticality~\cite{kom}, screening of superconducting fluctuations near the BCS-BEC crossover~\cite{luca}, etc.

It is, in principle, clear that the appearance of multiple characteristic lengths and the existence of many excitation gaps are both consequences of multiple sheets of the Fermi surface, interpreted as separate single-particle bands. However, whether or not one consequence implies the other remains unclear. In the present work we address this question and demonstrate that multi-condensate physics can take place irrespective of the presence/absence of multiple spectral gaps in the uniform superconducting state. These features appear on different levels of the theory - the system can have multiple energy gaps in the excitation spectrum but a single characteristic length and, vice versa, a multiband superconductor can have multiple coherence lengths but a single excitation gap. In particular, we find that the crossover between superconductivity types I and II - the intertype (IT) regime - is strongly affected by difference between healing lengths of multiple contributing condensates even when the corresponding excitation gaps are degenerate and cannot be distinguished. Our analysis is done using the formalism of the extended Ginzburg-Landau (EGL) theory~\cite{extGL3,extGL4} generalized to the case of an arbitrary number of contributing bands in this work. 

The paper is organized as follows. In Sec.~\ref{sec2} we discuss our formalism based on the $\tau$-expansion of the microscopic equations, with $\tau=1-T/T_c$ the proximity to the critical temperature. It goes to one order beyond the standard Ginzburg-Landau (GL) approach, which is sufficient to describe a finite IT domain between types I and II in the phase diagram of the superconducting magnetic response. This formalism is then used in Sec.~\ref{sec3}, where boundaries of the IT domain are obtained for different configurations of the multiband structure. Conclusions are given in Sec.~\ref{sec4}.

\section{Multiband EGL formalism}
\label{sec2}

The EGL formalism is a convenient tool that can be employed when the physics beyond the GL theory is of interest but full microscopic calculations are impractical. A relevant example is the crossover between superconductivity types I and II - the IT regime. It is well known that within the GL theory, the crossover is reduced  to a single point - it takes place at the critical GL parameter~\cite{landau,degen,kett} $\kappa =\kappa_0 =1/\sqrt{2}$~($\kappa = \lambda_L/\xi_{GL}$, where  $\lambda_L$ and $\xi_{GL}$ are the London magnetic penetration depth and GL coherence length). However, as is known since 70s, this GL-based picture is valid only in the limit $T \to T_c$~(more precisely, in the lowest order in $\tau$).  At  $T< T_c$~(beyond the lowest order in $\tau$) there is a finite temperature-dependent crossover interval of $\kappa$'s~\cite{krag,ess,aston,jac1,auer,klein, web,brandt1,luk,lav1,lav2,muhlb1, brandt2, pau, muhlb2,zir1,zir2}, which the GL theory does not capture. In the corresponding finite domain in the $\kappa$-$T$ plain (the IT domain),  the system has nonstandard field dependence of the magnetization~\cite{krag,ess,aston,jac1,auer,klein} with unconventional spatial configurations of the mixed state~\cite{extGL1,ess,lav1,lav2,muhlb1,brandt2,pau,muhlb2,zir1,zir2}, governed by long-range attraction of vortices~\cite{krag,ess,aston,jac1,auer,klein,web,brandt1,luk} and many-vortex interactions~\cite{extGL4a} - the so-called intermediate mixed state).

For the derivation of the EGL formalism, we employ the $M$-band generalization of the two-band BCS model~\cite{suhl,mos} with the $s$-wave pairing in all contributing bands and the Josephson-like Cooper-pair transfer between the bands. For illustration, we consider a system in the clean limit and assume that all available bands have parabolic single-particle energy dispersions with 3D spherical Fermi surfaces. The pairing is controlled by the symmetric real coupling matrix $\check{g}$, with the elements $g_{\nu\nu'}$. The derivation of the formalism comprises two main steps: (1) the multiband Neumann-Tewordt (NT) functional is obtained from the microscopic model and (2) the $\tau$-expansion is applied to reconstruct the NT functional. We outline main details of these steps, highlighting important differences in comparison with the two-band EGL approach~\cite{extGL4}. The obtained formalism is then used in the analysis of the boundaries of the IT domain in the $\kappa$-$T$ phase diagram. 

\subsection{Multiband Neumann-Tewordt functional}

The NT functional~\cite{nt1, nt2} is obtained from the microscopic expression for the condensate free energy by accounting for higher powers and higher gradients of the band gap functions $\Delta_{\nu}=\Delta_{\nu}({\bf x})$, as compared to the GL functional. Only the terms giving the GL theory and its leading corrections are taken into account. The general expression for the free energy density of $M$-band $s$-wave superconductor (relative to that of the normal state at zero field) can be written as~\cite{extGL1,extGL4}
\begin{align}
f  = \frac{{\bf B}^2}{8\pi} +\langle\vec{\Delta}^{\dagger}, \check{g}^{-1} \vec{\Delta}\rangle
+ \sum\limits_{\nu=1}^M f_{\nu}[\Delta_{\nu}],
\label{F}
\end{align}
where $\vec{\Delta}^{\dagger}=(\Delta^\ast_1, \Delta^\ast_2, \ldots,\Delta^\ast_M)$ and ${\bf B}$ is the magnetic field, $\langle \vec{a}^{\,\dagger},\vec{b}\rangle=\sum_\nu a^{\ast}_\nu b_\nu$ denotes the scalar product of vectors $\vec{a}$ and $\vec{b}$ in the band space, and the functional $f_{\nu}[\Delta_{\nu}]$ reads
\begin{align}
f_{\nu}=&-\sum\limits_{n=0}^{\infty}\frac{1}{n+1}\int\prod_{j=1}^{2n+1}d^3{\bf y}_j \,
K_{\nu,2n+1}({\bf x},\{{\bf y}\}_{2n+1})\notag\\
&\times \Delta_{\nu}^{\ast}({\bf x})\Delta_{\nu}({\bf y}_1)\ldots \Delta^{\ast}_{\nu}
({\bf y}_{2n})\Delta_{\nu}({\bf y}_{2n+1}),
\label{Fcal}
\end{align}
with $\{ {\bf y}\}_{2n+1}=\{{\bf y}_1, \ldots, {\bf y}_{2n+1}\}$. The integral kernels in Eq.~(\ref{Fcal}) are given by ($m$ is odd)
\begin{align}
K_{\nu,m}({\bf x},\{{\bf y}\}_m)=
&-T\,\sum\limits_{\omega}{\cal G}^{(B)}_{\nu,\omega}({\bf x},{\bf y}_1){\bar{\cal G}}^{(B)}_{\nu,\omega}
({\bf y}_1,{\bf y}_2)\ldots\notag \\
&\times{\cal G}^{(B)}_{\nu,\omega} ({\bf y}_{m-1},{\bf
y}_m){\bar{\cal G}}^{(B)}_{\nu,\omega}({\bf y}_m,{\bf x}),
\label{Knu}
\end{align}
where $\omega$ is the fermionic Matsubara frequency, ${\cal G}^{(B)}_{\nu,\omega}({\bf x},{\bf y})$ is the Fourier transform of the single-particle Green function calculated in the presence of the magnetic field and $\bar{\cal G}^{(B)}_{\nu,\omega}({\bf x}, {\bf y})=-{\cal G}^{(B)}_{\nu,-\omega}({\bf y},{\bf x})$. For ${\cal G}^{(B)}_{\nu,\omega}$ we employ the standard approximation sufficient to derive the extended GL theory
\begin{equation}
{\cal G}_{\nu,\omega}^{(B)}({\bf x},{\bf y}) = \exp\left[\mathbbm{i}
\,\frac{e}{\hbar \mathbbm{c}}\int_{{\bf y}}^{{\bf x}} {\bf A}({\bf z})\cdot
d{\bf z}\right] {\cal G}^{(0)}_{\nu,\omega}({\bf x},{\bf y}),
\label{GnuB}
\end{equation}
where the integral in the exponent is taken along the classical trajectory of a charge carrier in a magnetic field with the vector potential ${\bf A}$. Here the Green function for zero field writes
\begin{equation}
{\cal G}^{(0)}_{\nu,\omega}({\bf x},{\bf y})=\int\frac{d^3{\bf k}
}{(2\pi)^3}\frac{\exp[\mathbbm{i}{\bf k}\cdot({\bf x} - {\bf
y})]}{\mathbbm{i}\hbar\omega -\xi_{\nu}({\bf k})},
\label{Gnu}
\end{equation}
where the band-dependent single-particle energy dispersion reads
\begin{align}
\xi_{\nu}({\bf k})=\xi_{\nu}(0) + \frac{\hbar^2{\bf k}^2}{2m_{\nu}}-\mu,
\label{xi}
\end{align}
with $m_{\nu}$ the band effective mass, $\xi_{\nu}(0)$ the band lower edge, and $\mu$ the chemical potential.

To get simpler differential structure of the functional (\ref{F}), one invokes the gradient expansion for the band gap functions and the vector potential as 
\begin{align}
&\Delta_{\nu}({\bf y})  = \Delta_{\nu}({\bf x}) +\big(({\bf y} - {\bf x})\cdot 
\boldsymbol{\nabla}_{\bf x}\big) \Delta_{\nu}({\bf x})+\ldots ,\notag\\
&{\bf A}({\bf y})  = {\bf A}({\bf x}) +\big(({\bf y} - {\bf x})\cdot 
\boldsymbol{\nabla}_{\bf x}\big) {\bf A}({\bf x})+\ldots ,
\end{align} 
which makes it possible to represent non-local integrals in $f_{\nu}$ as a series in powers of $\Delta_{\nu}$, its gradients and field spatial derivatives. The series are infinite and therefore a truncation procedure is needed. The GL theory follows from the standard Gor'kov truncation~\cite{fett}. To incorporate the leading corrections to the GL formalism, one needs to go beyond the Gor'kov approximation. As the form of $f_{\nu}$ is not sensitive to the number of contributing bands, one can apply the truncation procedure to each of the band contributions separately and utilize the previous results derived for the single- and two-band cases, see Ref.~\onlinecite{extGL4}. The resulting multiband NT functional reads
\begin{align}
f =&\frac{{\bf B}^2}{8\pi}+\langle\vec{\Delta}^{\dagger}, \check{g}^{-1} \vec{\Delta}\rangle
+\sum\limits_{\nu=1}^M\biggl\{ \Big[{\cal A}_{\nu}+a_{\nu}\Big(\tau +\frac{\tau^2}{2} \Big)\Big]|\Delta_{\nu}|^2\notag\\
&+\frac{b_{\nu}}{2}  (1 + 2\tau )  |\Delta_{\nu}|^4-\frac{c_{\nu}}{3}|\Delta_{\nu}|^6
+ {\cal K}_{\nu} (1 + 2\tau) |{\bf D}\Delta_{\nu}|^2\notag\\
&- {\cal Q}_{\nu} \Big(|{\bf D}^2\Delta_{\nu}|^2+\frac{1}{3}{\rm rot}{\bf B}\cdot{\bf i}_{\nu}
+\frac{4e^2}{\hbar^2\mathbbm{c}^2}{\bf B}^2 |\Delta_{\nu}|^2 \Big)\notag\\
& - \frac{{\cal L}_{\nu} }{2}\Big[8|\Delta_{\nu}|^2|{\bf D}\Delta_{\nu}|^2+ \big(\Delta_{\nu}^{\ast 2}
({\bf D}\Delta_{\nu})^2+{\rm c.c.}\big)\Big]\biggr\},
\label{NT}
\end{align}
with ${\bf D}=\boldsymbol{\nabla}-\mathbbm{i} (2e/\hbar\mathbbm{c}){\bf A}$ and ${\bf i}_{\nu} =(4e/\hbar\,\mathbbm{c})\, {\rm Im}\big[\Delta_{\nu}^* {\bf D}\Delta_{\nu}\big]$. The band dependent coefficients in Eq.~(\ref{NT}) are
\begin{align}
&{\cal A}_\nu = N_\nu \ln\Big(\frac{2e^\Gamma\hbar \omega_c}{\pi
T_c}\Big), a_{\nu} = - N_{\nu}, b_{\nu}=\,N_{\nu}\frac{7\zeta(3)}{8\pi^2T_c^2},\notag\\
&c_{\nu}=N_{\nu}\,\frac{93\zeta(5)}{128\pi^4T_c^4},
{\cal K}_{\nu} = \frac{b_{\nu}}{6}\hbar^2 v_{\nu}^2, \; {\cal Q}_{\nu}=\frac{c_{\nu}}{30}
\hbar^4 v_{\nu}^4, \notag\\
&{\cal L}_{\nu} = \frac{c_{\nu}}{9}\hbar^2 v_{\nu}^2,
\label{coefficients}
\end{align}
where $\omega_c$ is the cut-off frequency, $N_{\nu}$ is the band DOS,  $v_{\nu}$ denotes the band Fermi velocity,  $T_c$ is in the energy units, and $\zeta(\ldots)$ and $\Gamma$ are the Riemann zeta-function and Euler constant.

The NT functional appears as a natural extension of the GL theory. The initial motivation of its derivation was to have an approach beyond the GL theory, which preserves, to some extent, the simplicity of the GL formalism. Such an approach is especially important in the case of spatially nonuniform problems. Unfortunately, the stationary point equations derived from the NT functional are rather complex even for the single-band case and not easier to solve than the original microscopic equations [see, e.g., Eq.~(3) in Ref.~\onlinecite{jac2}]. Furthermore, these equations admit unphysical solutions~\cite{jac2} such as weakly damped oscillations of the condensate near the core of a single vortex state. The roots of this problem lie in the fact that the NT free energy functional is not bound from below because the coefficients $c_{\nu}$, ${\cal Q}_{\nu}$, and ${\cal L}_{\nu}$ are positive. We also note in passing that a similar functional is commonly used in the analysis of the Fulde-Ferrel-Larkin-Ovchinnikov (FFLO) pairing [see, e.g., Refs.~\onlinecite{buz,buz1,min})], however, in that case the sign of $c_{\nu}$, ${\cal Q}_{\nu}$ and ${\cal L}_{\nu}$ is changed due the spin-magnetic interaction, which marks the appearance of the stable FFLO regime. 

\subsection{Perturbative $\tau$-expansion}
\label{sec:tau}

It was suggested for the single-band case (see Ref.~\onlinecite{jac2} and references therein) that to eliminate the nonphysical solutions, the Neumann-Tewordt functional should be restructured by applying the perturbative $\tau$-expansion, based on the fact that the fundamental small parameter of the microscopic equations is the proximity to the critical temperature $\tau$. The stationary solution for the order parameter within the Neumann-Tewordt approach contains all odd powers of $\tau^{1/2}$ while the truncation of the infinite series in Eq.~(\ref{Knu}) does not distort only the two lowest orders $\tau^{1/2}$~(the GL term) and $\tau^{3/2}$~(the leading correction to the GL term). Incomplete higher-order terms in $\tau$ should be removed by means of the $\tau$-expansion. A similar approach was subsequently applied to the two-band NT functional~\cite{extGL4}. Here we generalize it to the case of an arbitrary number of contributing bands $M$.

Following this approach, we represent the band gap functions and fields in the form of $\tau$-series~\cite{extGL3,extGL4} 
\begin{align}
&\Delta_{\nu}=\tau^{1/2} \big[\Delta^{(0)}_{\nu} + \tau\Delta^{(1)}_{\nu} + \ldots\big],\notag\\
&{\bf A}=\tau^{1/2} \big[ {\bf A}^{(0)} + \tau {\bf A}^{(1)} + \ldots\big],\notag\\
&{\bf B}=\tau \big[ {\bf B}^{(0)} + \tau {\bf B}^{(1)}+ \ldots\big].
\label{tauexp}
\end{align}
One also takes into account divergent condensate and field characteristic lengths $\propto \tau^{-1/2}$ that affect spatial gradients in the NT functional. This is formally done by introducing the spatial scaling as ${\bf x} \to \tau^{1/2}{\bf x}$ [see discussion after Eq.~(10) in Ref.~\onlinecite{extGL3}]. Notice that to get the stationary solution in the two lowest orders in $\tau$, one also needs to operate with $\Delta^{(2)}_{\nu}$ but only in intermediate expressions.   
   
Inserting Eq.~(\ref{tauexp}) into Eq.~(\ref{NT}) and applying the scaling ${\bf x} \to \tau^{1/2}{\bf x}$, one obtains the free energy density as
\begin{align}
f=\tau^2 \big[ \tau^{-1} f^{(-1)}+f^{(0)} + \tau f^{(1)}+ \ldots\big].
\label{ftau}
\end{align}
Notice that the two lowest orders in the band gap functions and the field produce three lowest orders in the free energy but, as is shown below, the contribution $f^{(-1)}$ is zero for the stationary point. This contribution reads as
\begin{align}
f^{(-1)} =\langle\vec{\Delta}^{(0)\dagger}, \check{L} \vec{\Delta}^{(0)}\rangle,
\label{f-1}
\end{align}
where the matrix elements of $\check{L}$ are defined as 
\begin{align}
L_{\nu\nu'}=g^{-1}_{\nu\nu'}-{\cal A}_{\nu} \delta_{\nu\nu'},
\label{checkL}
\end{align}
with $g^{-1}_{\nu\nu'}$ being elements of the inverse coupling matrix $\check{g}^{-1}$ and $\delta_{\nu\nu'}$ denoting the Kronecker symbol. The next-order term $f^{(0)}$ is the GL functional
\begin{align}
f^{(0)} =&\frac{{\bf B}^{(0)2}}{8\pi}+\Big(\langle\vec{\Delta}^{(0)\dagger}, \check{L} \vec{\Delta}^{(1)}\rangle
+ {\rm c.c.}\Big) + \sum\limits_{\nu=1}^Mf^{(0)}_{\nu},
\label{f0}
\end{align}
where $f^{(0)}_{\nu}$ is given by
\begin{align}
f^{(0)}_{\nu}=a_{\nu}|\Delta^{(0)}_{\nu}|^2+\frac{b_{\nu}}{2} |\Delta^{(0)}_{\nu}|^4
+ {\cal K}_{\nu} |{\bf D}^{(0)}\Delta^{(0)}_{\nu}|^2,
\label{f0nu}
\end{align}
and ${\bf D}^{(0)}=\boldsymbol{\nabla}-\mathbbm{i} (2e/\hbar\mathbbm{c}){\bf A}^{(0)}$. Finally, 
the highest-order term in Eq.~(\ref{ftau}) is given by
\begin{align}
f^{(1)} =&\frac{{\bf B}^{(0)}\cdot {\bf B}^{(1)}}{4\pi}+\Big(\langle\vec{\Delta}^{(0)\dagger}, \check{L} 
\vec{\Delta}^{(2)}\rangle+ {\rm c.c.}\Big)\notag\\
& + \langle\vec{\Delta}^{(1)\dagger}, \check{L} \vec{\Delta}^{(1)}\rangle+\sum\limits_{\nu=1}^M f^{(1)}_{\nu},
\label{f1}
\end{align}
where
\begin{align}
f^{(1)}_{\nu} =&\Big(a_{\nu}+b_{\nu}|\Delta^{(0)}_{\nu}|^2\Big) \Big(\Delta^{(0)*}_{\nu} \Delta^{(1)}_{\nu}
+ {\rm c.c.}\Big)+ \frac{a_{\nu}}{2}|\Delta^{(0)}_{\nu}|^2\notag\\
&+b_{\nu}|\Delta^{(0)}_{\nu}|^4-\frac{c_{\nu}}{3}|\Delta^{(0)}_{\nu}|^6
+ 2{\cal K}_{\nu}|{\bf D}^{(0)}\Delta^{(0)}_{\nu}|^2\notag\\
&+{\cal K}_{\nu}\Big[ \big( {\bf D}^{(0)} \Delta^{(0)}_{\nu}\cdot {\bf D}^{(0)\ast}\Delta^{(1)\ast}_{\nu}
+{\rm c.c.}\big)-{\bf A}^{(1)} \cdot {\bf i}_{\nu}^{(0)}\Big]\notag\\
&- {\cal Q}_{\nu} \Big(|{\bf D}^{(0)2}\Delta^{(0)}_{\nu}|^2
+\frac{1}{3}{\rm rot}{\bf B}^{(0)}\cdot {\bf i}^{(0)}_{\nu}+\frac{4e^2{\bf B}^{(0)2}}{\hbar^2\mathbbm{c}^2}
\notag\\
&\times|\Delta^{(0)}_{\nu}|^2 \Big) - \frac{{\cal L}_{\nu} }{2}\Big\{8|\Delta^{(0)}_{\nu}|^2| {\bf D}^{(0)} \Delta^{(0)}_{\nu}|^2\notag\\
&+ \big[\Delta^{(0)2}_{\nu}({\bf D}^{(0)\ast}\Delta^{(0)\ast}_{\nu})^2+{\rm c.c.}\big]\Big\},
\label{f1nu}
\end{align}
and ${\bf i}^{(0)}_{\nu}$ is the lowest-order term in the $\tau$-expansion of ${\bf i}_{\nu}$.

The $\tau$-expansion of the NT functional is then used to derive a set of the stationary-point equations for the gap functions and fields contributions - each of the equations correspond to a particular order of the $\tau$-expansion. The equation in the lowest order reads
\begin{align}
\frac{\delta{\cal F}^{(-1)}}{\delta \vec{\Delta}^{(0)\dagger}}=\check{L}\vec{\Delta}^{(0)}=0,
\label{st-1}
\end{align}
where $ {\cal F}^{(-1)}$ the free energy contribution obtained by integrating $f^{(-1)}$. This is the linearized gap equation in the multiband BCS theory that determines $T_c$. It has a nontrivial solution when 
\begin{align}
{\rm det} \check{L} =0.
\label{tc}
\end{align}
Recalling the definition of $\check{L}$, which includes ${\cal A}_{\nu}$ and, hence, depends on $T_c$, one sees that Eq.~(\ref{tc}) determines zeros of an $M$-degree polynomial of the variable $\ln(2e^\Gamma\hbar \omega_c/\pi T_c)$. One should choose the smallest root of this polynomial, which gives the largest $T_c$. Here we assume that this solution is non-degenerate. This implies that the solution of the gap equation Eq.~(\ref{st-1}) corresponds to a one-dimensional irreducible representation of the system symmetry group. The opposite occurs in a particular case when the superconducting system has a symmetry additional to $U(1)$, which is reflected in a special symmetry of the matrix $\check{L}$ and results in the appearance of multi-component order parameter [see details in Refs.~\onlinecite{extGL5} and \onlinecite{sigr}].

Once $T_c$ is determined, it is convenient to introduce the eigenvalues and eigenvectors of $\check{L}$ as
\begin{align}
\check{L}\vec{\epsilon} =0
\label{eps}
\end{align}
with the zero eigenvalue and 
\begin{align}
\check{L}\vec{\eta}_i =\Lambda_i \vec{\eta}_i,
\label{eta}
\end{align}
with nonzero eigenvalues $\Lambda_i\not=0$. As the matrix $\check{L}$ is real and symmetric, the vectors $\vec{\epsilon}$ and $\vec{\eta}_i$ can be chosen such that they form an orthonormal basis so that $\langle \vec{\,\epsilon}^{\dagger}, \vec{\epsilon} \rangle =1$, $\langle \vec{\,\epsilon}^{\dagger}, \vec{\eta}_i \rangle =0$ and $\langle \vec{\,\eta}_i^{\dagger}, \vec{\eta}_j \rangle =\delta_{ij}$. Then a general solution to the gap equation (\ref{st-1}) reads in the form
\begin{align}
\vec{\Delta}^{(0)}= \psi({\bf x}) \vec{\epsilon},
\label{Psi}
\end{align}
where $\psi({\bf x})$ controls the spatial profiles of all band condensates in the lowest order in $\tau$. 

The shape of $\psi({\bf x})$ is governed by the stationary point equations associated with the GL functional (\ref{f0}). The first one of those is given by
\begin{align}
\frac{\delta {\cal F}^{(0)}}{\delta \vec{\Delta}^{(0)\dagger}}=\check{L}\vec{\Delta}^{(1)}
+ \vec{W}[\vec{\Delta}^{(0)}]=0,
\label{st0_D}
\end{align}
where $ {\cal F}^{(0)}$ is the free-energy term corresponding to $f^{(0)}$ and the components of $\vec{W}$ read
\begin{align}
W_{\nu}=a_{\nu}\Delta^{(0)}_{\nu}+b_{\nu}\Delta^{(0)}|\Delta^{(0)}_{\nu}|^2
- {\cal K}_{\nu} {\bf D}^{(0)2}\Delta^{(0)}_{\nu}.
\label{wnu}
\end{align}
The second (Maxwell) equation is obtained as 
\begin{align}
\frac{\delta {\cal F}^{(0)}}{\delta {\bf A}^{(0)}}=\frac{1}{4\pi} {\rm rot}{\bf B}^{(0)}-\sum\limits_{\nu=1}^M {\cal K}_{\nu}{\bf i}^{(0)}_{\nu}=0.
\label{st0_A}
\end{align}
Notice that the equation $\delta {\cal F}^{(0)}/\delta \vec{\Delta}^{(1)\dagger}=0$ coincides with Eq. ~(\ref{st-1}) while  $\delta {\cal F}^{(0)}/\delta {\bf A}^{(1)\dagger}=0$ is an identity relation because ${\bf A}^{(1)}$ does not contribute to $f^{(0)}$. By projecting Eq.~(\ref{st0_D}) onto $\vec{\epsilon}$ and keeping in mind that $\vec{\,\epsilon}^{\dagger} \check{L}=0$, one gets  
\begin{align}
a\psi + b\psi|\psi|^2 - {\cal K}{\bf D}^{(0)2}\psi =0,
\label{GL1}
\end{align}
where coefficients $a$, $b$ and ${\cal K}$ are averages over the contributing bands
\begin{align}
a=\sum\limits_{\nu=1}^M a_{\nu} |\epsilon_{\nu}|^2, b=\sum\limits_{\nu=1}^M b_{\nu} |\epsilon_{\nu}|^4,
{\cal K}=\sum\limits_{\nu=1}^M {\cal K}_{\nu} |\epsilon_{\nu}|^2,
\label{abK}
\end{align}
and $\epsilon_{\nu} $ are the components of $\vec{\epsilon}$. Similarly, Eq.~(\ref{st0_A}) is reduced to
\begin{align}
{\rm rot}{\bf B}^{(0)}= 4\pi {\cal K}{\bf i}^{(0)}_{\psi},
\label{GL2}
\end{align}
where ${\bf i}^{(0)}_{\psi}$ is obtained from ${\bf i}^{(0)}_{\nu}$ by substituting $\psi$ for $\Delta^{(0)}_{\nu}$. 

Therefore, the GL equations for the $M$-band system are given by Eqs.~(\ref{GL1}) and (\ref{GL2}). The corresponding condensate state is described by a single-component order parameter $\psi({\bf x})$, in full agreement with the Landau theory in the case of a non-degenerate solution for $T_c$, see also Refs.~\onlinecite{kosh,kres,kog}. We note that the number of the components of the order parameter is determined by the dimensionality of the relevant irreducible representation of the corresponding symmetry group~\cite{landau}, not by the number of the bands. The single-component order parameter means that the standard classification of the superconducting magnetic response is applied here: we have types I and II with the IT regime in between. The presence of multiple bands is reflected only in the expressions for the coefficients $a$, $b$ and ${\cal K}$.

Using the eigenvectors of $\check{L}$ as the basis, we represent the next-to-leading contribution to the gap function as
\begin{align}
\vec{\Delta}^{(1)} = \varphi({\bf x}) \vec{\epsilon} + \sum\limits_{i=1}^{M-1} \varphi_i({\bf x}) \vec{\eta}_i,
\label{PhiPhi_i}
\end{align}
with new position-dependent functions $\varphi$ and $\varphi_i$ to be found. Inserting Eq.~(\ref{PhiPhi_i}) in Eq.~(\ref{st0_D}), one obtains the equation
\begin{align}
\sum\limits_{i=1}^{M-1} \Lambda_i \varphi_i \vec{\eta}_i + \vec{W}[\vec{\Delta}^{(0)}]=0.
\label{st0exp}
\end{align}
Equation~(\ref{st0exp}) is solved by projecting it onto $\vec{\eta}_j$, which yields $M-1$ equations for $\varphi_j $, i.e.,
\begin{align}
\varphi_j = -\frac{1}{\Lambda_j} \big( \alpha_j\psi + \beta_j\psi|\psi|^2 - \Gamma_j {\bf D}^{(0)2}\psi\big),
\label{GL3}
\end{align}
where the coefficients $\alpha_j,\beta_j$, and $\Gamma_j$ are of the form
\begin{align}
&\alpha_j=\sum\limits_{\nu=1}^M a_{\nu} \eta^*_{j\nu} \epsilon_{\nu},\;
 \beta_j=\sum\limits_{\nu=1}^M b_{\nu} \eta^*_{j\nu} \epsilon_{\nu} |\epsilon_{\nu}|^2,\notag\\
&\Gamma_j=\sum\limits_{\nu=1}^M {\cal K}_{\nu} \eta^*_{j\nu} \epsilon_{\nu}, 
\label{albetGam}
\end{align}
and $\eta_{j\nu} $ are components of $\vec{\eta}_j$. Equations~(\ref{PhiPhi_i}), (\ref{GL3}), and (\ref{albetGam}) generalize the corresponding expressions for the two-band case~\cite{extGL4}. One should keep in mind that the present formalism involves the eigenvectors of $\check{L}$ while $\vec{\Delta}^{(1)}$ in Ref.~\onlinecite{extGL4} was represented as a linear combination of other explicitly chosen vectors. Therefore, to recover the expression for $\vec{\Delta}^{(1)}$ in Ref.~\onlinecite{extGL4}, one needs to express $\vec{\epsilon}$ and $\vec{\eta}_j$ for $M=2$ in terms of the vectors used in Ref.~\onlinecite{extGL4}. 

Thus, $M-1$ functions $\varphi_i$, which determine the second term for $\vec{\Delta}^{(1)}$ in Eq.~(\ref{PhiPhi_i}), are found from the simple algebraic expressions (\ref{GL3}) when using solutions to the GL equations (\ref{GL1}) and (\ref{GL2}). To find the first term in Eq.~(\ref{PhiPhi_i}), that depends on $\varphi$ and the leading correction to the field ${\bf A}^{(1)}$, one needs to solve the system of equations resulting from the projection of Eq.~(\ref{st0exp}) onto the eigenvector $\vec\epsilon$ and zero functional derivatives of the free-energy contribution corresponding to $f^{(1)}$. However, as will be shown below, $\varphi$ and ${\bf A}^{(1)}$ do not contribute to the boundaries of the IT domain. We note, however, that $\varphi$ is necessary to calculate the band healing lengths - this calculation is outlined in the Appendix.

\subsection{Free energy at the stationary point and thermodynamic critical field}

The stationary free energy density is found by substituting the obtained stationary solutions into the corresponding expressions for the free energy functional, i.e., 
\begin{align}
f_{\rm st} = \tau^2 \big[f_{\rm st}^{(0)}+\tau f_{\rm st}^{(1)} + \ldots\big],
\label{ftaust}
\end{align}
where the term of the order $\tau$ is absent by the virtue of Eq.~(\ref{st-1}) and the first non-vanishing contribution is the GL free enegy
\begin{align}
f^{(0)}_{\rm st}= \frac{{\bf B}^{(0)2}}{8\pi} + a|\psi|^2+\frac{b}{2}|\psi|^4 + {\cal K}|{\bf D}^{(0)}\psi|^2,
\label{f0st}
\end{align}
We have also taken into account that $\vec{\Delta}^{(0)\dagger} \check{L} \vec{\Delta}^{(1)}=0$, which follows from Eq.~(\ref{st-1}). 

To find the leading order correction to the stationary GL free energy, we first rearrange the terms in $f^{(1)}$ that include $\Delta^{(1)}_{\nu} $ and $\Delta^{(1)\ast}_{\nu}$. For the stationary solution the sum of these terms in Eqs.~(\ref{f1}) can be represented  as
\begin{align}
\langle\vec{\Delta}^{(1)\dagger}, \check{L} \vec{\Delta}^{(1)}\rangle + \Big(\langle\vec{\Delta}^{(1)\dagger}, \vec{W}\rangle + {\rm c.c.}\Big)=- \langle\vec{\Delta}^{(1)\dagger},\check{L} \vec{\Delta}^{(1)}\rangle,
\end{align}
where Eq.~(\ref{st0_D}) is taken into consideration. Using Eqs.~(\ref{GL1}) and (\ref{GL3}), we further obtain that
$\langle\vec{\Delta}^{(1)\dagger}, \check{L} \vec{\Delta}^{(1)}\rangle$ can be expressed only in terms of $\psi$ as
\begin{align}
\langle\vec{\Delta}^{(1)\dagger}, \check{L} \vec{\Delta}^{(1)}\rangle =&|\psi|^2 \sum\limits_{i=1}^{M-1} \frac{a^2 |\bar{\alpha}_i|^2}{\Lambda_i} + 2|\psi|^4\sum\limits_{i=1}^{M-1} \frac{a b {\rm Re}[\bar{\alpha}^{\ast}_i \bar{\beta}_i]}{\Lambda_i} \notag\\
&+|\psi|^6 \sum\limits_{i=1}^{M-1} \frac{b^2 |\bar{\beta}_i|^2}{\Lambda_i},
\end{align}
where the dimensionless parameters $\bar{\alpha}_i$ and $\bar{\beta}_i$ are defined by
\begin{align}
\bar{\alpha}_i=\frac{\alpha_i}{a}-\frac{\Gamma_i}{{\cal K}}, \;
\bar{\beta}_i=\frac{\beta_i}{b}-\frac{\Gamma_i}{{\cal K}}.
\label{bar-ab}
\end{align}
Then, $f^{(1)}$, given by Eqs.~(\ref{f1}) and (\ref{f1nu}), can be represented for the stationary solution in the form
\begin{align}
f_{\rm st}^{(1)} =&\,\frac{{\bf B}^{(0)}\cdot {\bf B}^{(1)}\! -\! {\bf A}^{(1)}  \cdot {\rm rot}{\bf B}^{(0)}}{4\pi}+ \frac{\gamma_a}{2}|\psi|^2 +\gamma_b |\psi|^4\notag\\
&-\frac{\gamma_c}{3}|\psi|^6 + 2{\cal K}|{\bf D}^{(0)}\psi|^2- {\cal Q}\Big(|{\bf D}^{(0)2}\psi|^2\notag\\
&+\frac{1}{3}{\rm rot}{\bf B}^{(0)}\cdot {\bf i}^{(0)}_{\psi}+\frac{4e^2{\bf B}^{(0)2}}{\hbar^2\mathbbm{c}^2}|\psi|^2 \Big) \notag\\
&- \frac{{\cal L}}{2} \Big\{8|\psi|^2| {\bf D}^{(0)} \psi|^2+ {\rm Re}\big[\psi^2\big({\bf D}^{(0)\ast}\psi^\ast\big)^2\big]\Big\},
\label{f1st}
\end{align}
where
\begin{align}
{\cal Q}=\sum\limits_{\nu=1}^M {\cal Q}_{\nu} |\epsilon_{\nu}|^2, {\cal L}=\sum\limits_{\nu=1}^M {\cal L}_{\nu} |\epsilon_{\nu}|^4, c=\sum\limits_{\nu=1}^M c_{\nu} |\epsilon_{\nu}|^6.
\label{QLc}
\end{align}
and 
\begin{align}
& \gamma_a = a-2\sum\limits_{i=1}^{M-1} \frac{a^2 |\bar{\alpha}_i|^2}{\Lambda_i}, \,  \gamma_b = b-2\sum\limits_{i=1}^{M-1} \frac{a b {\rm Re}[\bar{\alpha}^{\ast}_i \bar{\beta}_i]}{\Lambda_i}, \notag \\
&  \gamma_c = c+3\sum\limits_{i=1}^{M-1} \frac{b^2 |\bar{\beta}_i|^2}{\Lambda_i}.
\label{QLgamma}
\end{align}

Using the above result, one can calculate the thermodynamic critical field $H_c$ which is also sought in the form of the $\tau$-expansion 
\begin{align}
H_c = \tau \big[H^{(0)}_c + \tau H^{(1)}_c + \ldots\big].
\label{tauexpHc}
\end{align} 
By virtue of the definition, $H_c$ can be obtained from
\begin{align}
\frac{H^2_c}{8\pi} = -f_{{\rm st},0},
\label{Hcdef}
\end{align}
where $f_{{\rm st},0}$ is the free energy density of the Meissner state. In the lowest (GL) order, the uniform solution of Eq.~(\ref{GL1}) is given by $\psi_0=\sqrt{|a|/b}$. This yields the corresponding contribution to the thermodynamic critical field as
\begin{align}
H^{(0)}_c=\sqrt{\frac{4\pi a^2}{b}},
\label{Hc0}
\end{align}
see Eqs.~(\ref{ftaust}), (\ref{f0st}), and (\ref{tauexpHc}). $H^{(0)}_c$ is formally the same as that for the single- and two-band cases~\cite{extGL3,extGL4} but with the difference that $a$ and $b$ are now averages over $M$ contributing bands. The next-order contribution to $H_c$ is obtained from Eqs.~(\ref{ftaust}), (\ref{f1st}), and (\ref{tauexpHc}), which gives
\begin{align}
\frac{H^{(1)}_c}{H^{(0)}_c} =-\frac{1}{2} - \frac{ca}{3b^2} - \sum\limits_{i=1}^{M-1} \frac{a}{\Lambda_i}
|\bar{\alpha}_i - \bar{\beta}_i|^2.
\label{Hc1}
\end{align}
Here the third term in the left-hand side has a different form as compared to the corresponding expressions for the single- and two-band cases in Refs.~\onlinecite{extGL3} and \onlinecite{extGL4}. The origin of the differences has been already discussed after Eq.~(\ref{albetGam}).

\subsection{Gibbs free energy difference}

A type I superconductor can only have a spatially uniform Meissner condensate state, which undergoes an abrupt transition to the normal state when the amplitude of the applied field ${\bf H}$ exceeds $H_c$. Type II superconductors, in addition to the Meissner phase, develop a non-uniform (mixed) state between the lower $H_{c1}$ and upper $H_{c2}$ critical fields, where $H_{c1} \leq H_c \leq H_{c2}$. A formal criterion for switching from type I to type II is that at $H=H_c$ the Meissner state becomes less energetically favourable than the mixed state. It is investigated by using the Gibbs free energy so that the switching criterion is obtained as the vanishing difference between the Gibbs free energies of the Meissner and non-uniform states. The Gibbs free energy density for a superconductor at $H=H_c$ is given by $g=f_{\rm st} -H_c B/4\pi$, where ${\bf B}$ is directed along the external field ${\bf H}$ and found from the stationary-point equations for the corresponding condensate state. For the Meissner state we have $B=0$ and $g =g_0=f_{{\rm st},0}=-H^2_c/8\pi$. Thus, the density of the Gibbs free energy difference $\g= g-g_0$ is written as
\begin{align}
\g=f_{\rm st} -\frac{H_c B}{4\pi}+\frac{H^2_c}{8\pi}.
\label{dg}
\end{align}

To calculate $\g$, it is convenient to the use dimensionless quantities
\begin{align}
&\tilde{\bf x}=\frac{\bf x}{\sqrt{2}\lambda}, \; \tilde{\bf A} = \kappa \frac{\bf A}{\lambda_L H^{(0)}_c},\; \tilde{\bf B} = \sqrt{2}\kappa \frac{\bf B}{H^{(0)}_c},\notag\\
&\tilde{\psi} = \frac{\psi}{\psi_0}, \; \tilde{\g} =\frac{4\pi \g}{H^{(0)2}},\;
\tilde{G}=\frac{4\pi G}{H^{(0)2} (\sqrt{2}\lambda_L)^3} , 
\label{dim}
\end{align}
where $G$ is the integral of $\g$ and
\begin{align}
\lambda_L = \frac{\hbar \mathbbm{c}}{|e|}\sqrt{\frac{b}{32\pi K|a|}}, \;
\kappa =\frac{\lambda_L}{\xi_{GL}}= \lambda_L \sqrt{\frac{|a|}{K}}.
\label{ch-lengths}
\end{align}
We note that Eq.~(\ref{ch-lengths}) differs from the conventional definitions for the GL coherence length $\xi_{GL}$ and London penetration depth $\lambda_L$ by the absence of the factor $\tau^{-1/2}$. This difference appears due to the scaling ${\bf x} \to \tau^{1/2}{\bf x}$ used in the derivation of the $\tau$-expansion. Using the dimensionless units, we write the GL equations as
\begin{align}
\psi -\psi|\psi|^2 +\frac{1}{2\kappa^2} {\bf D}^{(0)2}\psi  =0,\;
{\rm rot}\, {\bf B}^{(0)}= {\bf i}^{(0)}_{\psi},
\label{GLscale}
\end{align}
${\bf D}^{(0)}= \boldsymbol{\nabla} + \mathbbm{i}{\bf A}^{(0)}$,  ${\bf i}^{(0)}_{\psi}=2{\rm Im}\big[\psi {\bf D}^{(0)\ast}\psi^{\ast}\big]$ and the spatial gradients are also dimensionless. Hereafter we omit the tilde for brevity.

The $\tau$-expansion for $\g$ is obtained from Eqs.~(\ref{ftaust}), (\ref{f0st}), (\ref{f1st}), and (\ref{dg}) in the form
\begin{align}
\g = \tau^2 \big[\g^{(0)} + \tau \g^{(1)} + \ldots\big],
\label{dgtau}
\end{align}
where
\begin{align}
\g^{(0)} =&\frac{1}{2}\bigg(1-\frac{B^{(0)}}{\sqrt{2}\kappa}\bigg)^2 - |\psi|^2 +\frac{1}{2}|\psi|^4
+\frac{|{\bf D}^{(0)}\psi|^2}{2\kappa^2}
\label{g0}
\end{align}
and  
\begin{align}
\g^{(1)} = &\left(1-\frac{B^{(0)}}{\sqrt{2}\kappa}\right) \left(\frac{\bar{\gamma}_a}{2}-\bar{c}\bar{\gamma}_c -\bar{\gamma}_b \right)-\frac{\bar{\gamma}_a}{2} |\psi|^2 + \bar{\gamma}_b  |\psi|^4\notag\\
& + \bar{c} \bar{\gamma}_c  |\psi|^6 +\frac{|{\bf D}^{(0)}\psi|^2}{\kappa^2} +\frac{\bar{\cal Q}}{4\kappa^4}\Big(|{\bf D}^{(0)2}\psi|^2\notag\\
&+\frac{{\bf i}^{(0)2}_{\psi}}{3}+{\bf B}^{(0)2}|\psi|^2 \Big)+\frac{\bar{\cal L}}{4\kappa^2} \Big\{8|\psi|^2|{\bf D}^{(0)} \psi|^2\notag\\
&+ {\rm Re}\big[\psi^2\big({\bf D}^{(0)\ast}\psi^\ast\big)^2\big]\Big\},
\label{g1}
\end{align}
where the dimensionless parameters are given by
\begin{align}
&\bar{c}=\frac{ca}{3b^2},\; \bar{\cal Q}=\frac{{\cal Q}a}{{\cal K}^2}, \;
\bar{\cal L}=\frac{{\cal L}a}{{\cal K}b},  \;  \bar{\Lambda}_j=\frac{\Lambda_j}{a}, \notag \\
&\bar \gamma_a=\frac{\gamma_a}{a},\; \bar \gamma_b=\frac{\gamma_b}{b}, \;
\bar \gamma_c=\frac{\gamma_c}{c}.
\label{bar-par}
\end{align}

We note that to derive Eq.~(\ref{g1}), we rearrange Eq.~(\ref{f1st}) by using the identity ${\bf A}^{(1)} \cdot {\rm rot}{\bf B}^{(0)}={\bf A}^{(1)} \cdot {\rm rot}({\bf B}^{(0)}-{\bf H}^{(0)}_c)$. It is then integrated by parts, giving ${\bf B}^{(1)} \cdot ({\bf B}^{(0)}-{\bf H}^{(0)}_c )$ while surface integrals vanish. This makes sure that the next-to-lowest order contribution to ${\bf B}$ does not appear in the Gibbs free energy difference, similarly to $\varphi$ contributing to $\vec{\Delta}^{(1)}$. Thus, the Gibbs free energy difference, taken in the lowest and next-to-lowest orders in $\tau$ depends only on the solution to the GL equations. One also notes that the GL contribution $\g^{(0)}$ is not sensitive to $M$ which enters only its leading correction $\g^{(1)}$.

\subsection{B point and intertype domain}

Integrating Eq.~(\ref{dgtau}) yields the Gibbs free energy difference $G$. However, since the goal of our study is superconducting properties in the vicinity of the B point, in addition to the $\tau$-expansion we apply the expansion with respect to $\delta \kappa = \kappa - \kappa_0$, which gives
\begin{align}
\frac{G}{\tau^{1/2}} = &G^{(0)}  +\frac{dG^{(0)}}{d\kappa} \delta\kappa + G^{(1)}\tau, 
\label{dgtaukap}
\end{align}
where only the linear contributions in $\propto \delta\kappa$ and $\propto \tau$ are kept and  the expansion coefficients are calculated at $\kappa = \kappa_0$. A significant advantage of this approach is that at $\kappa_0$ the GL theory simplifies considerably because the condensate-field configurations become self-dual being related by~\cite{bog}
\begin{equation}
B^{(0)}=1- |\Psi|^2,
\label{bog1}
\end{equation}
while the order parameter $\psi$ satisfies the first order differential equation
\begin{equation}
\left( D^{(0)}_x - \mathbbm{i} D^{(0)}_y \right) \psi=0.
\label{bog2}
\end{equation}
Here the field is taken along the $z$-direction so that $\psi$ is not dependent on $z$ and one can use ${\bf D}^{(0)2}=D^{(0)2}_x + D^{(0)2}_y$. Equations (\ref{bog1}) and (\ref{bog2}) are often referred to as the Bogomolnyi equations~\cite{bog} (in the context of superconductivity they are also known as the Sarma solution~\cite{degen}).  
Using these equations, one can demonstrate that the first contribution to the Gibbs free energy difference $G^{(0)}$ vanishes identically for any solution of the GL equations, which is a manifestation of the fact that at $H=H_c$ the self-dual GL theory is infinitely degenerate. The GL theory predicts that at $\kappa_0$ the normal state $\psi=0$ is stable above $H_c$ while below $H_c$ the Meissner state $\psi=1$ appears. Then, the mixed state appears only at $H=H_c$, hosting a plethora of exotic condensate-field configurations. Corrections to the GL theory break the degeneracy and successive self-dual configurations of the magnetic flux and condensate determine the properties of the IT mixed state. 

With the Bogomolnyi equations, the Gibbs free energy difference given by Eq.~(\ref{dgtaukap}) is reduced to
\begin{align}
\frac{G}{\tau^{1/2} L}= -\sqrt{2}\,{\cal I} \,\delta\kappa +  ({\cal I} {\cal A}+ {\cal J}{\cal B}) \, \tau,
\label{GJI}
\end{align}
where $L$ is the system size along the direction of the field,  ${\cal I}$ and ${\cal J} $ are given by the integrals 
\begin{align}
&{\cal I} =  \!\! \int d^3{\bf x}|\psi|^2 \big(1 - |\psi|^2\big), {\cal
J} =\!\! \int d^3{\bf x}|\psi|^4 \big(1 - |\psi|^2\big),
\label{IJ}
\end{align}
while coefficients ${\cal A}$ and ${\cal B}$ are given by
\begin{align}
&{\cal A}=2(1+\bar{\cal Q}) -\bar{\gamma}_b -\bar{c}\bar{\gamma}_c,\;
{\cal B} = 2\bar{\cal L}-\frac{5}{3}\bar{\cal Q} - \bar{c}\bar{\gamma}_c.
\label{calAB}
\end{align}
Apart from the constants, that depend on $M$ contributing bands, this expression for the Gibbs free energy difference is the same as obtained earlier for single- and two-band superconductors~\cite{extGL1}. 

Now we have everything at our disposal to determine the boundaries of the IT domain on the $\kappa$-$T$ plane. Its lower boundary $\kappa^{\ast}_{\rm min}(T)$ separates type I and IT regimes and marks the appearance/disappearance of the mixed state~\cite{extGL1}. At this boundary the upper critical field $H_{c2}$ approaches $H_c$. The condensate vanishes at $H_{c2}$ and so the Gibbs free energies of the normal and condensate states become equal. At the same time the normal and Meissner states have the same Gibbs free energy at $H_c$. Therefore, the lower boundary of the IT domain is found from the criterion $G = 0$ taken together with the condition $\psi \to 0$. The latter means ${\cal J}/{\cal I} =0$ in Eq.~(\ref{GJI}). Then one finds 
\begin{align}
\kappa^\ast_{\rm min}=\kappa_0 (1 + \tau {\cal A}).
\label{kmin}
\end{align} 

The upper boundary $\kappa^{\ast}_{\rm max}(T)$ separates type II and IT regimes and is determined by changing the sign of the long range interaction between vortices~\cite{extGL1} - it is repulsive in type II and attractive in the IT domain. In order to calculate $\kappa^{\ast}_{\rm max}(T)$, one finds the asymptote of the GL solution for two vortices at large distance between them. The position dependent part of this asymptotic solution is plugged into Eq.~(\ref{GJI}), which yields the long-range interaction potential between two vortices. As the scaled GL equations (\ref{GLscale}) are independent of the number of contributing bands, one can adopt the long-range asymptote of the two-vortex solution $\psi$ found previously in the two-band case~\cite{extGL1}, which yields ${\cal J}/{\cal I}=2$. Then, the upper boundary is obtained as
\begin{align}
\kappa^\ast_{\rm max}=\kappa_0 \big[ 1 + \tau\big( {\cal A}+2{\cal B}\big) \big].
\label{kmax}
\end{align} 

\section{Role of multiple bands}
\label{sec3}

\subsection{General observations}

A transparent structure of all contributions in the EGL formalism makes it possible to obtain important preliminary results before calculating $\kappa^{\ast}_{\rm min}$ and $\kappa^{\ast}_{\rm max}$. The most significant observation is that the multigap structure and the disparity between characteristic lengths of different band condensates appear on different levels of the theory, leading to different physical consequences. Multiple excitation gaps appear in the lowest order in $\tau$ of the EGL theory: following Eq.~(\ref{Psi}), a multiband superconductor in the GL regime has, in general, multiple excitation gaps while the contributing band condensates are governed by the unique GL coherence length $\xi_{GL}$. Thus, on the level of the GL theory superconducting magnetic properties of a multiband system are the same as those of the single-band superconductor having the only energy gap in the excitation spectrum. 
 
Differences between the condensate characteristic lengths and, thus, between spatial profiles of different band condensates appear only when corrections to the GL theory are taken into account, i.e., in the next-to-lowest order in $\tau$. Using the EGL approach, one calculates the band condensate healing lengths, to find a band-dependent leading correction to $\xi_{GL}$ as $|\xi_\nu - \xi_{\nu^\prime}| \propto \tau \xi_{GL}$~[see Appendix~\ref{appB} and Ref.~\onlinecite{extGL6}]. Thus, one can expect that phenomena associated with the disparity between the band condensate lengths are notable only at sufficiently low temperatures. 

However, an important exception is the vicinity of the B point, i.e., the IT domain between types I and II. Here the GL theory is close to degeneracy and the next-to-lowest corrections in $\tau$~(and, thus, the difference between the band condensate lengths) play a crucial role in shaping the superconducting magnetic properties. In this case the mixed state becomes very sensitive to all characteristics of the multiband system, including the number of contributing bands and parameters of multiple Fermi sheets comprising the complex Fermi surface. The multiband structure can, therefore, have a notable effect on the IT domain, justifying the focus of this work.

It is also of significance, that the number of the energy gaps in the excitation spectrum of a uniform multiband superconductor is not always equal to the number of the contributing bands $M$, which can be seen from the corresponding gap equation
\begin{align}
\Delta_{\nu}=\sum\limits_{\nu'=1}^M \lambda_{\nu\nu'} n_{\nu'}\!\!\int\limits_{0}^{\hbar\omega_c} \!d\varepsilon\,\frac{\Delta_{\nu'}}{E_{\nu'}} \big[1-2f(E_{\nu'})\big],
\label{multigaps}
\end{align}
where $E_{\nu}=\sqrt{\varepsilon^2 + |\Delta_{\nu}|^2}$ is the single-particle excitation energy, $\lambda_{\nu \nu'}=g_{\nu\nu'} N$ denotes the dimensionless coupling constant, $N=\sum_{\nu}N_{\nu}$  is the total single-particle density of states (DOS),  $n_{\nu}=N_{\nu}/N$ is the relative DOS for band $\nu$,  $f(E_{\nu})$ is the Fermi distribution function, and $\omega_c$ is the cut-off frequency. The excitation gaps become degenerate when the quantity 
\begin{align}
D_\nu = \sum\limits_{\nu'=1}^M \lambda^{}_{\nu\nu'}n^{}_{\nu'}
\end{align}
assumes the same value for several bands. 

\subsection{Microscopic parameters}
\label{sec:mat_const}

\begin{figure*}[t]
\includegraphics[width=1.0\linewidth]{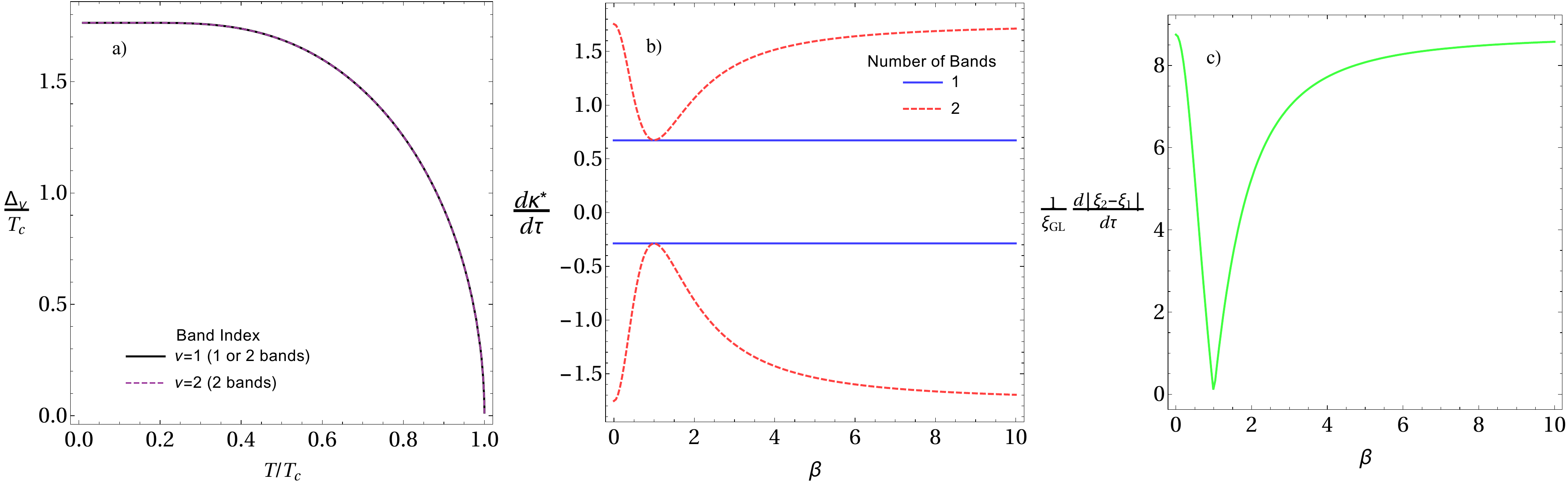}
\caption{Results for $1$- and $2$-band superconductors with the microscopic parameters given in Sec. \ref{sec:mat_const}, chosen so that both materials have the same excitation gap, degenerate for the 2-band system, shown in panel a) as a function of $T$~(in units of $T_c$). Panel b) plots slopes of the IT domain boundaries $d\kappa_{\rm max}^{\ast}/d\tau$~(two upper lines) and  $d\kappa_{\rm min}^{\ast}/d\tau$~(two lower lines) versus $\beta=v_2/v_1$ for the two-band (dotted) and single-band (solid) cases, the single-band results are material independent quantities $-0.29$ and $0.67$. Panel c) shows the derivative $d|\xi_2-\xi_1|/d\tau$ versus $\beta$, where $|\xi_2-\xi_1|$ is the absolute value of the difference of the band healings lengths $\xi_2$ and $\xi_1$ for the two-band system in question.}
\label{fig1}
\end{figure*}

The IT domain boundaries $\kappa^{\ast}_{\rm min}$ and $\kappa^{\ast}_{\rm max}$ depend on the following microscopic parameters: the dimensionless couplings $\lambda_{\nu\nu'}=g_{\nu\nu'} N$~(with $N=\sum_{\nu}N_{\nu}$ the total DOS), the relative band DOSs $n_{\nu}=N_{\nu}/N$, and the band velocities ratios $v_{\nu}/v_1$. Since $T_c \propto \hbar\omega_c$, the cut-off frequency $\omega_c$ does not contribute to $\kappa^{\ast}_{\rm min}$ and $\kappa^{\ast}_{\rm max}$. 

For the calculations we choose realistic values of the parameters, recalling that in two-band superconductors the intraband dimensionless couplings are typically in the range $0.2$-$0.7$ while the interband coupling is much smaller  [see Ref.~\onlinecite{extGL1} and references therein]. The relative band DOSs are usually similar for all bands. The range of $v_{\nu}/v_1$ can be estimated from the first principle calculations as well as from the ARPES measurements. For example, the angle-averaged Fermi velocities in the $a$-$b$ plane of ${\rm MgB}_2$ are calculated from first principles as $v_{\sigma}^{(a-b)}= 4.4 \times 10^5 m/s$ for the $\sigma$ states and $v_{\pi}^{(a-b)}= 5.35 \times 10^5 m/s$ for the $\pi$ states~\cite{brink}. However, for the $c$-direction such calculations yield $v_{\sigma}^{(c)}= 7 \times 10^4 m/s$ which is by an order of magnitude smaller than $v_{\pi}^{(c)}= 6 \times 10^5 m/s$.  In addition, ARPES measurements for iron chalcogenide ${\rm FeSe}_{0.35}{\rm Te}_{0.65}$ have revealed three contributing bands with the maximal ratio of the band Fermi velocities close to $4$ [see Ref.~\onlinecite{lub} and discussion in Ref.~\onlinecite{extGL6}]. 

In order to illustrate the role of the multiband structure in the presence of degenerate gaps, we compare results obtained for $1$- and $2$-band materials in Fig.~\ref{fig1} and for $2-$ and $4$-band superconductors in  Fig.~\ref{fig2}. For the first comparison we choose $\lambda=0.35$ for the $1$-band material and $\lambda_{11}=\lambda_{22}=0.3$ and $\lambda_{12}=0.05$ for the $2$-band system. This choice ensures that the both variants exhibit the same single energy gap in the excitation spectrum. To compare the $2$- and $4$-band materials, we take $\lambda_{11}=0.175$, $\lambda_{22}=0.125$, $\lambda_{12}=0.05$ and $n_1=n_2$ for the $2$-band system and  $\lambda_{11}=\lambda_{22}=0.3$, $\lambda_{33}=\lambda_{44}=0.2$ and $\lambda_{\nu \not=\nu'}=0.05$ for the $4$-band superconductor. These parameters are chosen to satisfy conditions $D_{\nu=1,2}^{(4)} = D_{\nu=1}^{(2)}$  and $D_{\nu=3,4}^{(4)} = D_{\nu=2}^{(2)}$, which gives the same two excitation gaps for both cases. To illustrate variations in the boundaries of the IT domain, we assume that the relative Fermi velocities depend on the variable parameter $\beta$: for $2$ band systems we take $v_2/v_1=\beta$ whereas for the $4$-band system we set the relations $\beta = v_2/v_1 = v_3/v_1 = 2 v_4/v_1$. The relative band DOSs for all contributing bands are assumed equal. 

We stress that our qualitative conclusions do not depend on a particular choice of the microscopic parameters.

\subsection{Numerical results for the IT domain boundaries}

\begin{figure*}[t]
\includegraphics[width=0.65\linewidth]{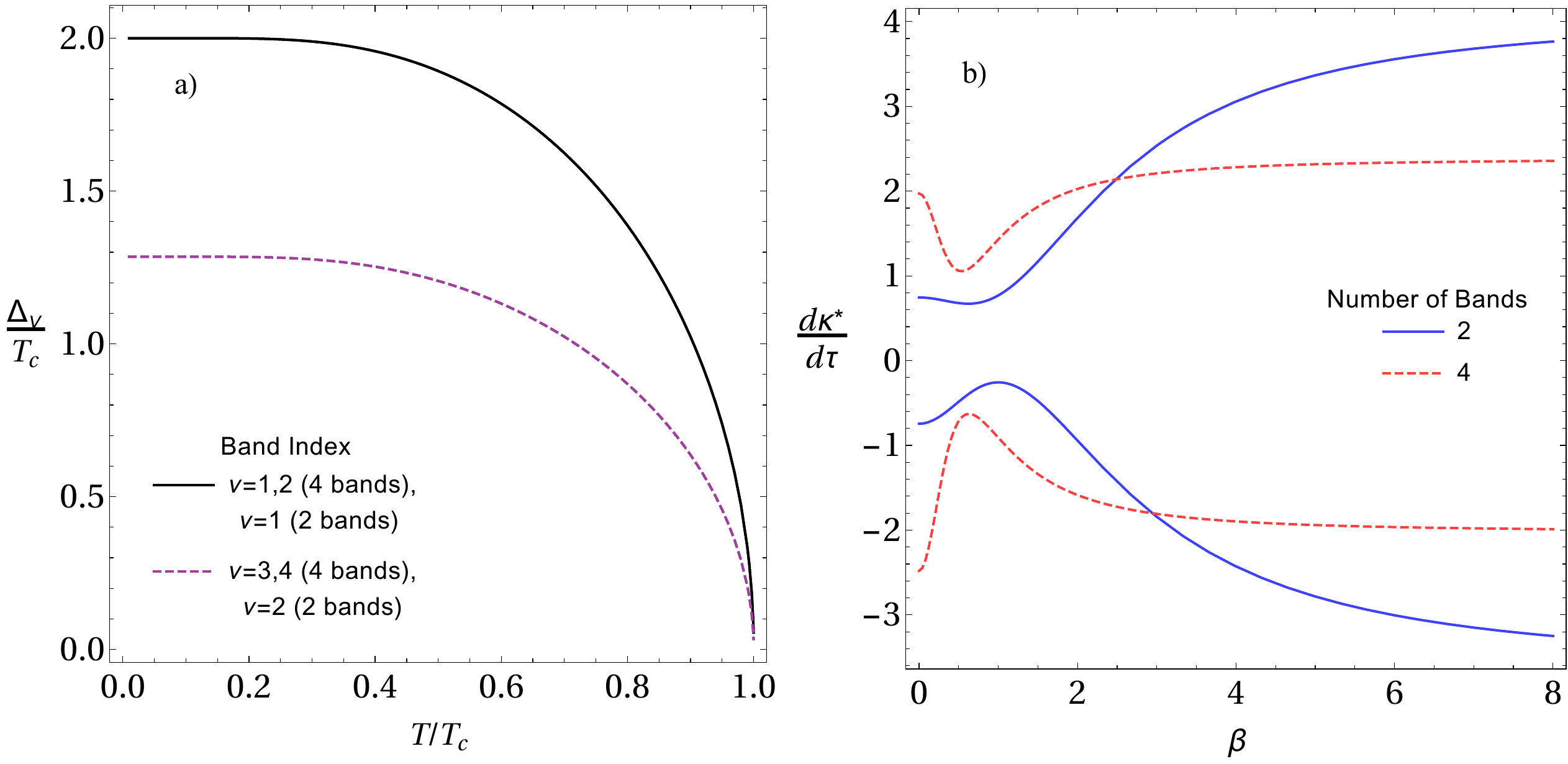}
\caption{Results for $2$- and $4$-band materials, calculated with the microscopic parameters given in Sec.~\ref{sec:mat_const} and chosen such that both materials have two excitation gaps. Panel a) shows the gaps versus temperature (in units of $T_c$). Panel b) plots slopes of the IT domain boundries $d\kappa_{\rm max}^{\ast}/d\tau$~(two upper lines) and  $d\kappa_{\rm min}^{\ast}/d\tau$~(two lower lines) as functions of $\beta=v_2/v_1$ (for the $2$-band case) and $\beta=v_2/v_1 = v_3/v_1=v_4/2v_1$ (for the $4$-band system).  }
\label{fig2}
\end{figure*}

Using the chosen microscopic parameters, we examine excitation gaps, the boundaries of the IT domain as well as the condensate healing lengths for superconductors with one, two and four bands. 

Figure~\ref{fig1} illustrates a comparison between the $1$- and $2$-band models. The both models exhibit the same single energy gap in the excitation spectrum, as shown in Fig.~\ref{fig1} a). However, the difference between them is apparent in Fig.~\ref{fig1} b) which shows $d\kappa^{\ast}_{\rm min}/d\tau$ and $d\kappa^{\ast}_{\rm max}/d\tau$ as functions of the Fermi velocities ratio $\beta =v_2/v_1$. The calculation reveals a notable dependence of the IT domain boundaries of the 2-band model on $\beta$~(dotted lines) in comparison with the $1$-band case, for which the IT boundaries are given by the material-independent constants $-0.29$ and $0.67$,~\cite{extGL1} as illustrated by solid lines. The difference between the two cases is maximal in the limits $\beta \ll 1$ and $\beta \gg 1$ but disappears when $\beta = 1$. 

To clarify the physical roots of the obtained results, we utilize the formalism of Ref.~\onlinecite{extGL6} (for reader's convenience, outlined in Appendix~\ref{appB}), and calculate the derivative $d|\xi_2-\xi_1|/d\tau$, where $|\xi_2-\xi_1|$ is the absolute value of the difference of the band healing lengths $\xi_2$ and $\xi_1$ for the two-band system in question. To the next-to-lowest order in $\tau$, we have $\xi_{\nu}=\xi^{(0)}_{\nu} + \tau \xi^{(1)}_{\nu}$, with $\xi^{(0)}_{\nu}=\xi_{GL}$, see Refs.~\onlinecite{extGL4, kres,kog,extGL6}. [This healing-length expression should be multiplied by $\tau^{-1/2}$ to return to the standard definition]. Therefore, taken to one order beyond the GL theory, $d|\xi_2-\xi_1|/d\tau$ is equal to $|\xi^{(1)}_2 -\xi^{(1)}_2|$ and is not $\tau$-dependent. The result is given in Fig.~\ref{fig1} c) in units of the GL coherence length $\xi_{GL}$. Comparing Figs.~\ref{fig1} b) and c) demonstrates that the size of the IT domain closely follows the healing length difference - the domain size grows with increasing the difference. One can thus see that even though the $2$-band system has a single gap in its excitation spectrum, its magnetic properties are strongly affected by the presence of multiple condensates with different characteristic lengths and, in general, differ significantly from those of the single-band case. The only exception is the case of $\xi_1 = \xi_2$, when the excitation spectra and magnetic properties of the single- and two-band systems become indistinguishable.

A further illustration is given in Fig.~\ref{fig2} which compares results for the $2$- and $4$-band systems. The parameters are chosen such that both systems have the same two excitation gaps [see Fig.~\ref{fig2} a)]. In particular, spectral gaps are degenerate for bands $1,2$ and $2,3$ in the $4$-band case. The IT domain boundaries (their $\tau$-derivatives) for the $2$-band system are shown in Fig.~\ref{fig2} b) versus $\beta=v_2/v_1$ by solid lines. One can see that in general, the corresponding IT domain is significantly different from the $2$-band IT domain shown in Fig.~\ref{fig1} b), which is a consequence of the two excitation gaps in the present case. However, the $2$-band IT boundaries in Fig.~\ref{fig2} b) are still close to the single-band ones in vicinity of $\beta=1$. Here the difference between the healing lengths $\xi_1$ and $\xi_2$ is minimal and the two-band system exhibits a nearly single-band superconducting magnetic response, despite the presence of two excitation gaps. We again observe that the presence/absence of diverse characteristic lengths of multiple condensates coexisting in one material is more essential for the superconducting magnetic properties than the presence/absence of multiple gaps in the excitation spectrum of the uniform superconductor. 

The quantities $d\kappa^{\ast}_{\rm max}/d\tau$ and $d\kappa^{\ast}_{\rm min}/d\tau$ for the four-band system are given by dotted lines in Fig.~\ref{fig2} b) versus $\beta=v_2/v1 = v_2/v_1 = 2 v_4/v_1$. One sees that the IT domain boundaries for the $4$-band case are close to the $2$-band IT boundaries at $\beta \sim 3$. For $\beta \gtrsim 3$ the size of the IT domain for the $2$-band system is notably larger and, on the contrary, for $\beta \lesssim 3$ the IT domain is larger in the $4$-band case. One also notes that unlike the $2$-band case the IT domain for the $4$-band system in Fig.~\ref{fig2} b) never approaches the single-band result [c.f. Fig.~\ref{fig1} b)]. In general, one can expect that the larger is the number of competing condensates, the more significant are the deviations from the single-condensate physics. 

\section{Conclusions}
\label{sec4}

In this work we have demonstrated that the presence of multiple competing lengths, each connected with corresponding partial condensate, is a more fundamental feature of a multiband superconductor for its magnetic properties than the presence of multiple gaps in the excitation spectrum. This is illustrated by considering boundaries of the IT domain in the phase diagram of the superconducting magnetic response. For example, our results have revealed that a superconductor can have many gaps in the excitation spectrum while exhibiting standard magnetic properties of a single-band material. There is also a reverse situation, when a superconductor has a single energy gap in the excitation spectrum but  multiple competing characteristic lengths of contributing band condensates, which results in notable changes of the superconducting magnetic properties in the IT regime as compared to the single-band case. Generally, our analysis shows that the multi-condensate physics can appear irrespective of the presence/absence of multiple spectral gaps. Two superconductors with different numbers of the contributing bands but with the same energy gaps in their excitation spectra (some of the spectral gaps are degenerate) can exhibit different magnetic properties sensitive to the spatial scales of the band condensates. This discrepancy between different manifestations of multiple bands in superconducting materials must be taken into account in analysis of experimental data and, generally, in studies of multiband superconductors. In addition, given the significant advances in chemical engineering of various materials, including multiband superconductors, it is of great importance to search for systems that  enrich our knowledge of and understanding   the physics of the materials. Multiband superconductors with degenerate excitation gaps can be a good example of such systems, clearly demonstrating that ``multiband" can be dramatically different from ``multigap".

Our analysis has been performed within the EGL approach that takes into account the leading corrections to the GL theory in the perturbative expansion of the microscopic equations in $\tau=1-T/T_c$. This formalism, previously constructed for single- and two-band systems, has been extended in the present work to the case of an arbitrary number of contributing bands. Its advantage is that it allows one to clearly distinguish various effects appearing due to the multiband structure in different types of superconducting characteristics. It particular, it reveals solid correlations between changes in the IT domain with the competition of multiple characteristic lengths of the contributing condensates.

\acknowledgements
This work was supported by Brazilian agencies, Conselho Nacional de Ci\^{e}ncia e Tecnologia (CNPq) , grant No. 309374/2016-2 and Funda\c{c}\~{a}o de Amparo a Ci\^encia e Tecnologia do Estado de Pernambuco (FACEPE) grant No. APQ-0936-1.05/15.  P. J. F. C. thanks CAPES Programa de Doutorado Sandu\'{i}che no Exterior, Processo No. 88881.186964/2018-01. T. T. S. and A. V. acknowledge hospitality of Departamento de F\'{i}sica da Universidade Federal de Pernambuco during their temporary stays in 2018 and support from CNPq grant No. 400510/2014-6 and FACEPE. M. D. C. acknowledges the Departamento de F\'{i}sica da Universidade Federal de Pernambuco for visiting professor fellowship in 2019, grant Propesq 05.2018- 031782/2018-88.

\appendix
\section{Leading correction to the GL coherence length}
\label{appB}

He we employ the EGL approach to calculate the band dependent healing lengths $\xi_{\nu}$ up to the leading corrections to the GL coherence length. The GL theory of multiband superconductors has a single order parameter which yields equal healing lengths for different band condensates. [A multiband superconductor can have more than one order parameter in the GL regime when the solution of the linearized gap equation for $T_c$ is degenerate~\cite{extGL5}; this case  is not considered here.] However, when one takes into account the leading corrections to GL theory, band healing lengths become different. These corrections have been calculated earlier~\cite{extGL6} for the $1$-band and $2$-band systems, and we now recall those results and extend them to the case of an arbitrary number of contributing bands. 

We consider the condensate that occupies a half space $x>0$ and is suppressed for $x \leq 0$. Each band condensate recovers its bulk value in a distance (measured from the interface $x=0$) that is called the band healing length $\xi_{\nu}$. This length is defined from the criterion
\begin{equation}
\frac{\Delta_{\nu}(\xi_{\nu})}{\Delta_{\nu}(\infty)} = \frac{\Delta_{\nu}^{(0)}(\xi_{\nu}^{(0)})}{\Delta_{\nu}^{(0)}(\infty)},
\end{equation}
where $\xi_{\nu}$ is given by the $\tau$-expansion
\begin{align}
\xi_{\nu}=\xi_{\nu}^{(0)}\big(1+\tau\xi_{\nu}^{(1)}\big),
\end{align}
and, taken in the lowest order in $\tau$, the band healing length coincides with the GL coherence length $\xi^{(0)}_{\nu}=\xi_{GL}$. We solve the GL equation~(\ref{GLscale}) without magnetic field and with the boundary conditions $\psi(0)=\psi'(\infty) =0$, with $\psi'$ the first derivative with respect to $x$ measured in units of $\xi_{GL}$. The well-known solution reads as~\cite{poole}
\begin{align}
\psi=\tanh\big(x/\sqrt{2}\big),
\label{Bpsitanh}
\end{align}
where $\psi$ is given in units of $\psi_0=\sqrt{|a|/b}$.  One sees from Eq.~(\ref{Psi}) that $\psi$ controls $\vec{\Delta}^{(0)}$. 

The next-to-lowest contribution to $\Delta_{\nu}$ is given by Eq.~(\ref{PhiPhi_i}). In this equation $\varphi_j$ are explicitly expressed via $\psi$ by Eq.~(\ref{GL3}) but $\varphi$ should be obtained from the stationary equation $\delta {\cal F}^{(1)}/\delta \vec{\Delta}^{(0)\dagger}= 0$, where $ {\cal F}^{(1)}$ corresponds to $f^{(1)}$ in Eq.~(\ref{ftau}). The projection of this equation onto the eigenvector $\vec\epsilon$ yields the equation for $\varphi$ that can be written as 
\begin{align}
(1-3\psi^2)\varphi+\varphi''=A\psi+B\psi^3+C\psi^5+D \psi {\psi'}^2,
\label{Bvarphipsi}
\end{align}
where $\varphi$ is in units of $\psi_0$, $\varphi''$ is the second derivative with respect to the scaled variable $x$, and the coefficients read as
\begin{align}
A=&\frac{3}{2}+\bar{\mathcal{Q}} +\sum_{i=1}^{M-1}\frac{\bar{\alpha}_i^2}{\bar{\Lambda}_i},\notag\\
B=&5\bar{\mathcal{L}} -4\bar{\mathcal{Q}}-2\sum_{i=1}^{M-1}\frac{\Gamma_i(\bar{\alpha}_i-\bar{\beta}_i)+2{\cal K}\bar{\alpha}_i\bar{\beta}_i}{\mathcal{K}\bar{\Lambda}_i},\notag\\
C=&3\bar{c}+3\bar{\mathcal{Q}} -5\bar{\mathcal{L}}+3\sum_{i=1}^{M-1} \frac{\bar{\beta}_i^2}{\bar{\Lambda}_i},\notag\\
D=&6\bar{\mathcal{Q}}-5\bar{\mathcal{L}}-6\sum_{i=1}^{M-1}\frac{\Gamma_i \bar{\beta}_i}{\mathcal{K}\bar{\Lambda}_i},
\end{align}
where ${\cal K}$, $\Gamma_j$, $\bar{c}$, $\bar{\cal Q}$, $\bar{\cal L}$, $\bar{\alpha}_j$, $\bar{\beta}_j$ and $\bar{\Lambda}_j$ are given by Eqs.~(\ref{abK}), (\ref{albetGam}), (\ref{bar-ab}), and (\ref{bar-par}). Here we consider that vectors $\vec\epsilon$ and $\vec\eta_j$ have only real components, i.e., $\alpha_i=\alpha^\ast_i$,
$\beta_i=\beta^\ast_i$, and $\Gamma_i=\Gamma^\ast_i$.

The solution of Eq.~(\ref{Bvarphipsi}) at $\varphi(0)=\varphi^\prime(\infty) =0$ is obtained as
\begin{align}
\varphi=&-\frac{3(A+B)+5C+D}{6}\tanh\left(\frac{x}{\sqrt{2}}\right)+\frac{2C+D}{6} \notag\\
&\times\tanh^3\left(\frac{x}{\sqrt{2}}\right)-\frac{A-C}{2}\frac{x}{\sqrt{2}}\;\mbox{sech}^2\left(\frac{x}{\sqrt{2}}\right).
\label{Bsolvarphi}
\end{align}
Then, using Eqs.~(\ref{Psi}), (\ref{PhiPhi_i}), (\ref{Bpsitanh}), and (\ref{Bsolvarphi}), one finds 
\begin{align}
\xi^{(1)}_{\nu}=&\frac{A-C}{2}+\sqrt{2}\psi\left(\frac{1}{\sqrt{2}}\right)\bigg(\frac{2C+D}{6}+\sum_{i=1}^{M-1}
\frac{\bar{\beta}_i}{\bar{\Lambda}_i}\frac{\eta_{i\nu}}{\epsilon_\nu}\bigg).
\label{Bxi1}
\end{align}
We note that only the last term in this expression contributes to the difference between the healing lengths of  two different bands $\nu$ and $\nu^\prime$, so that
\begin{align}
\xi_{\nu} - \xi_{\nu^\prime} = 0.86\, \tau\, \xi_{GL} \sum_{i=1}^{M-1} \frac{\bar{\beta}_i}{\bar{\Lambda}_i} \left(\frac{\eta_{i\nu}}{\epsilon_\nu} - \frac{\eta_{i\nu^\prime}}{\epsilon_{\nu^\prime}} \right),
\label{eq:diff_healing}
\end{align}
with $\psi(1/\sqrt{2})=0.86$. 

Let us consider, for illustration, Eq.~(\ref{Bxi1}) for the two-band system with degenerate excitation gaps. In this case the eigenvector of the matrix $\check{L}$ with zero eigenvalue satisfies $\epsilon_1=\epsilon_2$. Then, for equal band Fermi velocities the parameter $\bar{\beta}_1=0$ as $\beta_1/b = \Gamma_1/\cal K$, see the definition of these quantities in Sec.~\ref{sec:tau}. Since only $i=1$ contributes to the sum in Eq.~(\ref{Bxi1}), we find $\xi^{(1)}_1=\xi^{(1)}_2$. Thus, the healing lengths $\xi_1$ and $\xi_2$ are the same (at least up to the leading correction to the GL theory), which is in agreement with the results given in Fig.~\ref{fig1} c), where the healing length difference drops to zero at $v_2 = v_1$. The same conclusion can easily be obtained for $M>2$, when all excitation gaps are degenerate and the bands have the same Fermi velocity.


\end{document}